\documentclass[useAMS,usenatbib]{mn2e}

\usepackage{times}
\usepackage{amssymb}
\usepackage{graphicx}
\usepackage{graphics}

\def\asca{{\it ASCA\/}}

\def\chandra{{\it Chandra\/}}

\def\hst{{\it {\it HST}\/}}
\def\iras{{\it IRAS\/}}

\def\rosat{{\it ROSAT\/}}

\def\scuba{{\rm SCUBA\/}}

\def\xmm{{\it XMM-Newton\/}}

\def\aap{{A\&A}}
\def\aj{{AJ}}
\def\apj{{ApJ}}
\def\apjl{{ApJ}}
\def\apjs{{ApJS}}
\def\mnras{{MNRAS}}
\def\nat{{Nature}}

\def\etal{{et\,al.\,}}

\def\ltsima{$\; \buildrel < \over \sim \;$}
\def\simlt{\lower.5ex\hbox{\ltsima}}
\def\gtsima{$\; \buildrel > \over \sim \;$}
\def\simgt{\lower.5ex\hbox{\gtsima}}

\title[A Compton-thick quasar in FSC~10214+4724?]{A {\it Chandra} observation of the \boldmath{$z=2.285$} galaxy FSC~10214+4724: Evidence for a Compton-thick quasar?}
\author[D. M. Alexander et al.]
{D. M. Alexander,$^{1}$\thanks{E-mail: dma@ast.cam.ac.uk} G. Chartas,$^{2}$ F. E. Bauer,$^{1}$ W. N. Brandt,$^{2}$ C. Simpson,$^{3}$ and C. Vignali$^{4,5}$ \\ \\
$^{1}$Institute of Astronomy, Madingley Road, Cambridge CB3~0HA\\
$^{2}$Department of Astronomy \& Astrophysics, 525 Davey Laboratory, 
The Pennsylvania State University, University Park, PA 16802, USA\\
$^{3}$Department of Physics, University of Durham, South Road, Durham DH1 3LE\\
$^{4}$INAF - Osservatorio Astronomico di Bologna, Via Ranzani 1,
40127 Bologna, Italy\\
$^{5}$Dipartimento di Astronomia, Universit\'a degli Studi di Bologna, Via Ranzani 1, 40127 Bologna, Italy\\
}

\begin{document}

\date{Accepted ???. Received ???; in original form ???}

\pagerange{\pageref{firstpage}--\pageref{lastpage}} \pubyear{2004}

\maketitle

\label{firstpage}

\begin{abstract}
  
  We present a $\approx$~20~ks {\it Chandra} ACIS-S observation of the
  strongly lensed \hbox{$z=2.285$} ultra-luminous infrared galaxy
  FSC~10214+4724. Although this observation achieves the equivalent
  sensitivity of an up-to $\approx$~4~Ms {\it Chandra} exposure (when
  corrected for gravitational lensing), the rest-frame 1.6--26.3~keV
  emission from FSC~10214+4724 is weak ($L_{\rm
    X}\approx$~2$\times10^{42}$~erg~s$^{-1}$ for a lensing boost of
  $\approx100$); a significant fraction of this X-ray emission appears
  to be due to vigorous star-formation activity. If FSC~10214+4724
  hosts a quasar, as previously suggested, then it must be obscured by
  Compton-thick material. We compare FSC~10214+4724 to high-redshift
  \scuba\ galaxies and discuss the X-ray identification of
  Compton-thick AGNs at high redshift.

\end{abstract}


\section[]{Introduction}

The $z=2.285$ galaxy FSC~10214+4724 was one of the most remarkable
objects detected by the \iras\ survey. Originally proposed to be the
most luminous galaxy known (Rowan-Robinson \etal 1991),
multi-wavelength observations subsequently showed that it is lensed by
an intervening $z\approx0.9$ galaxy, boosting its intrinsic emission
by a factor of $\simgt$~10--100 (depending on the location and extent
of the unlensed emission with respect to the caustic; e.g.,\
Broadhurst \& Lehar 1995; Downes \etal 1995; Trentham 1995; Eisenhardt
\etal 1996; Evans \etal 1999). Optical and near-IR
spectroscopic/polarimetric observations have unambiguously shown that
FSC~10214+4724 hosts an obscured Active Galactic Nucleus (AGN; e.g.,\
Elston \etal 1994; Soifer \etal 1995; Goodrich \etal 1996).
Multi-wavelength analyses have suggested that the AGN is powerful
(e.g.,\ Goodrich \etal 1996; Granato \etal 1996; Green \&
Rowan-Robinson 1996), although it is generally accepted that
star-formation activity dominates the bolometric output (e.g.,\
Rowan-Robinson \etal 1993; Rowan-Robinson 2000).

The lensing-corrected properties of FSC~10214+4724 are similar to
those of \scuba\ galaxies (e.g.,\ Blain \etal 2002; Ivison \etal 2002;
Smail \etal 2002; Chapman \etal 2003; Neri \etal 2003): it lies at
$z>1$, is optically faint with $I\approx$~25, has an 850$\mu$m flux
density of a few mJy and a 1.4~GHz flux density of
$\approx$~25~$\mu$Jy, is massive (a molecular gas mass of $\approx
10^{10}$--$10^{11}$~$M_{\odot}$), and has a bolometric luminosity of
$\approx10^{13}$~L$_{\odot}$ (e.g.,\ Rowan-Robinson \etal 1993; Downes
\etal 1995; Eisenhardt \etal 1996). \scuba\ galaxies appear to be
massive galaxies undergoing intense star-formation activity and are
likely to be the progenitors of the local $\simgt$~$M_*$ early-type
galaxy population (e.g.,\ Blain \etal 2004; Chapman \etal 2004). Deep
X-ray observations have shown that a large fraction (at least
$\approx$~40\%; Alexander \etal 2003b, 2004) of the \scuba\ galaxy
population is actively fuelling its black holes during this period
of intense star formation, which can account for a significant
fraction of the black-hole growth in massive galaxies (Alexander \etal
2004).  The AGN in FSC~10214+4724 was identified via optical/near-IR
observations, and the large lensing boost of the AGN emission provides
interesting insight into AGN activity in the \scuba\ galaxy
population.

In this letter we present the results from a $\approx$~20~ks
observation of FSC~10214+4724 with the {\it Chandra X-ray Observatory}
(hereafter {\it Chandra}; Weisskopf et~al. 2000). Due to the large
lensing boost of FSC~10214+4724, these observations provide the
equivalent sensitivity of an up-to $\approx$~4~Ms \chandra\ exposure.
The Galactic column density toward FSC~10214+4724 is \hbox{1.2$\times
  10^{20}$~cm$^{-2}$} (Stark et~al.  1992).  $H_0=65$~km~s$^{-1}$
Mpc$^{-1}$, $\Omega_{\rm M}=1/3$, and $\Omega_{\Lambda}=2/3$ are
adopted.

%
%%%%%%%%%%%%%%%%%%%%%%%%%%%%%%%%%%%%%%%%%%%%%%%%%%%%%%%%%%%%%%%%%%%%%%
% FIGURE 1: Chandra-HST overlay
%%%%%%%%%%%%%%%%%%%%%%%%%%%%%%%%%%%%%%%%%%%%%%%%%%%%%%%%%%%%%%%%%%%%%%
%

\begin{figure*}
  \vspace*{-0.6in}
  \includegraphics[angle=0, width=150mm]{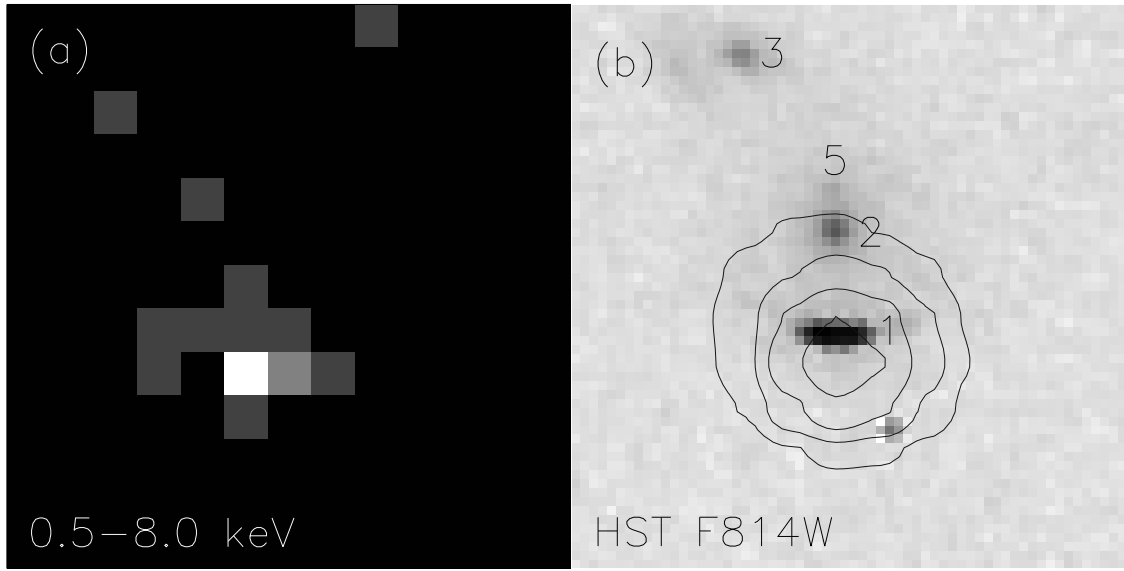}
  \vspace*{-0.6in}
  \caption{(a) Full-band \chandra\ image of FSC~10214+4724, and (b) F814W \hst\ image of FSC~10214+4724 with overlaid adaptively smoothed full-band contours. The \hst\ image was downloaded from the \hst\ archive and has been adjusted to give the same astrometry as that reported in Eisenhardt \etal (1996). The adaptive smoothing has been performed using the code of Ebeling, White, \& Rangarajan (2004) at the 2~$\sigma$ level; the contours are linear. Both images are $\approx$~$6\farcs 3\times6\farcs 3$; the full field of view of the ACIS S3 CCD is $8\farcm 6\times8\farcm 6$. Four of the components identified by Eisenhardt \etal (1996) are indicated; FSC~10214+4724 corresponds to component 1 and the $z\approx$~0.9 lensing galaxy corresponds to component 2. The X-ray detected source is clearly component 1.}
\end{figure*}

\section[]{{\it Chandra} Observation and Analysis}

FSC~10214+4724 was observed with {\it Chandra} on 2004 March 4
(observation ID 4807). The Advanced CCD Imaging Spectrometer (ACIS;
Garmire et~al. 2003) with the CCD S3 at the aim point was used for the
observation (the CCDs S1--S4, and I2--I3 were also turned
on);\footnote{For additional information on the ACIS and {\it Chandra}
  see the {\it Chandra} Proposers' Observatory Guide at
  http://cxc.harvard.edu/udocs/docs.} the optical position of
FSC~10214+4724 is $\alpha_{2000}=$~10$^{\rm h}$ 24$^{\rm m}$ 34\fs 54,
$\delta_{2000}=$~$+47^\circ$09$^{\prime}09\farcs8$. Faint mode was
used for the event telemetry format, and the data were initially
processed by the {\it Chandra} X-ray Center (CXC) using version 7.1.1
of the pipeline software.

%
%%%%%%%%%%%%%%%%%%%%%%%%%%%%%%%%%%%%%%%%%%%%%%%%%%%%%%%%%%%%%%%%%%%%%%
% TABLE 1: Chandra properties
%%%%%%%%%%%%%%%%%%%%%%%%%%%%%%%%%%%%%%%%%%%%%%%%%%%%%%%%%%%%%%%%%%%%%%
%

\begin{table*}
 \centering
  \caption{{\it Chandra} properties of FSC~10214+4724}
  \begin{tabular}{@{}ccccccccccc@{}}
  \hline
   \multicolumn{2}{c}{{\it Chandra} co-ordinates} & {\it Chandra}-{\it HST} & \multicolumn{3}{c}{Counts} & Band & Effective & \multicolumn{3}{c}{Flux} \\
   $\alpha_{2000}$ & $\delta_{2000}$ & (arcsec)$^{a}$ & FB$^{b}$ & SB$^{b}$ & HB$^{b}$ & Ratio$^{c}$ & $\Gamma$$^{d}$ & FB$^{e}$ & SB$^{e}$ & HB$^{e}$\\ 
 \hline
 10:24:34.560 & +47:09:09.48 & 0.17 & 13.6$^{+5.2}_{-3.3}$ & 9.9$^{+4.4}_{-3.0}$ & 3.8$^{+3.4}_{-1.7}$ & 0.38$^{+0.38}_{-0.21}$ & 1.6$^{+0.7}_{-0.6}$ & 5.4 & 2.0 & 3.6 \\
\hline
\end{tabular}
\begin{minipage}{170mm}
%\begin{tabular}{l}
Notes: $^a$Offset between the {\it Chandra} full-band source position and the {\it HST} source position (component 1) of Eisenhardt et~al. (1996). $^b$Source counts and errors. ``FB'' indicates full band, ``SB'' indicates soft band, and ``HB'' indicates hard band. The source counts are determined with {\sc wavdetect}. The errors correspond to 1~$\sigma$ and are taken from Gehrels (1986). $^c$Ratio of the count rates in the 2.0--8.0~keV and 0.5--2.0~keV bands. The errors were calculated following the ``numerical method" described in \S1.7.3 of Lyons (1991). $^d$Effective photon index for the 0.5--8.0~keV band, calculated from the band ratio. The photon index is related to the energy index by $\alpha$~=~$\Gamma-1$ where $F_{\nu}\propto\nu^{-\alpha}$. $^e$Fluxes in units of $10^{-15}$~erg~cm$^{-2}$~s$^{-1}$. These fluxes have been calculated from the count rate in each band using the CXC's Portable, Interactive, Multi-Mission Simulator (PIMMS) assuming  $\Gamma=1.6$; they have been corrected for Galactic absorption.\\
%\end{tabular}
\end{minipage}
\end{table*}

The reduction and analysis of the data used {\it Chandra} Interactive
Analysis of Observations ({\sc ciao}) Version~3.0.2
tools.\footnote{See http://cxc.harvard.edu/ciao/ for details on {\sc
    ciao}.} The {\sc ciao} tool {\sc acis\_process\_events} was used
to remove the standard pixel randomisation. The data were then
corrected for the radiation damage sustained by the CCDs during the
first few months of {\it Chandra} operations using the Charge Transfer
Inefficiency (CTI) correction procedure of Townsley et~al. (2000). All
bad columns, bad pixels, and cosmic ray afterglows were removed using
the ``status'' information in the event files, and we only used data
taken during times within the CXC-generated good-time intervals. The
background light curve was analysed to search for periods of
heightened background activity using the contributed {\sc ciao} tool
{\sc analyze\_ltcrv} with the data binned into 200~s intervals; there
were no periods of high background (i.e.,\ a factor of $\simgt$~2
above the median level). The net exposure time for the observation is
21.26~ks. The {\it ASCA} grade 0, 2, 3, 4, and 6 events were used in
all subsequent \chandra\ analyses.

\subsection{Source Identification}

The pointing accuracy of {\it Chandra} is excellent (the 90\%
uncertainty is $\approx0\farcs 6$).\footnote{See
  http://cxc.harvard.edu/cal/ASPECT/.} However, for our observation we
wanted to improve the source positions to provide an unambiguous
distinction between FSC~10214+4724 and nearby objects (e.g.,\ the
$z=0.9$ lensing galaxy lies $\approx$~$1\farcs2$ from FSC~10214+4724;
Eisenhardt et~al. 1996). To achieve this we matched sources detected
in the \chandra\ observation to sources found in the Sloan Digital Sky
Survey (SDSS), which has a positional accuracy of $\approx0\farcs 1$
(rms; Pier et~al. 2003).\footnote{Details about the SDSS and the SDSS
  catalogues can be accessed from http://www.sdss.org/.} \chandra\
source searching was performed using {\sc wavdetect} (Freeman \etal
2002) with a false-positive threshold of $1\times 10^{-5}$ in the full
(FB; 0.5--8.0~keV), soft (SB; 0.5--2.0~keV), and hard (HB; 2--8~keV)
bands; we used wavelet scale sizes of 1, 1.44, 2, 2.88, 4, 5.66, and 8
pixels.  The resulting source lists were then merged with a
$2^{\prime\prime}$ matching radius, producing a catalog of 37 sources.
The three brightest X-ray sources in this catalog (those with $>$20
counts in the full band) that lay within $5^{\prime}$ of the aim point
were matched to SDSS sources in the Data Release 2 catalog (DR2;
Abazajian \etal 2004) with a $2^{\prime\prime}$ search radius. The
mean SDSS-{\it Chandra} positional offset of the matched sources
(excluding FSC~10214+4724) was \hbox{--$0\farcs 18$$\pm0\farcs08$}
(right ascension; RA) and \hbox{--$0\farcs 28$$\pm0\farcs22$}
(declination; DEC).  These corrections were applied to the X-ray
source positions.

FSC~10214+4724 is detected in all three bands (see Table~1). The X-ray
source position (taken from the full band) lies $0\farcs 17$ from the
position measured by the {\it Hubble Space Telescope} (hereafter \hst)
in the F814W band (Eisenhardt et~al.  1996; see Table~1 \& Figure~1);
the X-ray position is also $0\farcs 38$ offset from both the CO(3--2)
position (Downes \etal 1995) and the 1.49~GHz radio position (Lawrence
\etal 1993). The X-ray source is clearly identified with
FSC~10214+4724 rather than the $z=0.9$ lensing galaxy (see Figure~1).

\subsection[]{Analysis}

The X-ray properties of FSC~10214+4724 are shown in Table~1. Although
this observation achieves the equivalent sensitivity of an up-to
$\approx$~4~Ms \chandra\ exposure (e.g.,\ a 10-count source has a
full-band flux of $\simgt$~$4\times10^{-17}$~erg~cm$^{-2}$~s$^{-1}$ for
a lensing boost of $\simlt100$; compare to Table~9 in Alexander \etal
2003a), only a few X-ray counts are detected. The $\approx$~10 counts
in the soft band correspond to a significant detection in the
rest-frame \hbox{1.6--6.6~keV} band while the $\approx$~4 counts in the hard
band correspond to a weak detection in the rest-frame 6.6--26.3~keV
band.  The soft-band flux is $\approx$~10 times below the 2~$\sigma$
\rosat\ constraint reported in Lawrence \etal (1994) and the
hard-band flux is $\approx$~20 times below the \asca\ upper limit
reported in Iwasawa (2001). We re-examined the \rosat\ PSPC image and
could not find unambiguous evidence of X-ray emission at the location
of FSC~10214+4724. From our analyses of the \rosat\ PSPC image we
determine a 3~$\sigma$ 0.5--2.0~keV upper limit of
$<1.3\times10^{-14}$~erg~cm$^{-2}$~s$^{-1}$. The observed (uncorrected
for gravitational lensing) full-band luminosity is $L_{\rm
  1.6-26.3~keV}=$~2.4~$\times10^{44}$~erg~s$^{-1}$.

%We do not find evidence for count-rate variability (the
%Kolmogorov-Smirnov test probability is 47\%; a 2-pixel radius aperture
%was used to extract the X-ray events)

The band ratio (i.e.,\ the ratio of the hard to soft-band count rate)
of FSC~10214+4724 implies an effective photon index of
$\Gamma=1.6^{+0.7}_{-0.6}$ (see Table~1). This is generally consistent
with that of an unobscured AGN (i.e.,\ $\Gamma\approx$~2.0; e.g.,\
Nandra \& Pounds 1994; George \etal 2000); however, due to the large
uncertainties and comparatively high redshift of FSC~10214+4724, this
could also be consistent with a column density of $N_{\rm H}\approx
2\times10^{23}$~cm$^{-2}$ at $z=2.285$ (for an intrinsic X-ray
spectral slope of $\Gamma=2.0$). The latter would be more consistent
with the obscured AGN classification of FSC~10214+4724 than the former
(e.g.,\ Elston \etal 1994; Soifer \etal 1995). See \S3.1 for further
obscuration constraints.

The gravitational lensing boost of FSC~10214+4724 is unknown in the
X-ray band. Since the lensing boost is a function of the source size,
basic constraints can be placed from the extent of the X-ray emission
(e.g.,\ see \S2 of Broadhurst \& Lehar 1995). In Figure~2 we show the
full-band profiles (S-N and E-W orientations) of FSC~10214+4724 and
compare them to the on-axis \chandra\ ACIS-S point spread function
(PSF). While this analysis is limited by small-number statistics, the
extent of FSC~10214+4724 is consistent with that of an unresolved
X-ray source ($\approx$~1$^{\prime\prime}$). The half-power radius of
FSC~10214+4724 (i.e.,\ the radius over which the central seven counts
are distributed; \hbox{$\approx$~0\farcs5--0\farcs75}) is also
consistent with that of an unresolved X-ray source. Although somewhat
uncertain, this suggests that the magnification in the X-ray band is
$\simgt$~25 (e.g.,\ compare to the extent and magnification found in
the 2.05~$\mu$m NICMOS observations of Evans \etal 1999). We can
compare this estimate to the expected lensing boost from other
observations.  The {\it HST} observations and source model of Nguyen
\etal (1999) suggest that the central source (i.e.,\ the \hbox{X-ray}
emitting AGN) is $\approx$~100~pc from the caustic, indicating that
the magnification of the AGN emission is likely to be $\approx$~100.
The strong optical polarisation and prominent high-excitation emission
lines also indicate that the caustic lies close to the central source
(e.g.,\ Broadhurst \& Lehar 1995; Lacy \etal 1998; Simpson \etal
2004).

%
%%%%%%%%%%%%%%%%%%%%%%%%%%%%%%%%%%%%%%%%%%%%%%%%%%%%%%%%%%%%%%%%%%%%%%
% FIGURE 2: X-ray extent
%%%%%%%%%%%%%%%%%%%%%%%%%%%%%%%%%%%%%%%%%%%%%%%%%%%%%%%%%%%%%%%%%%%%%%
%

\begin{figure}
  \includegraphics[angle=0,width=85mm]{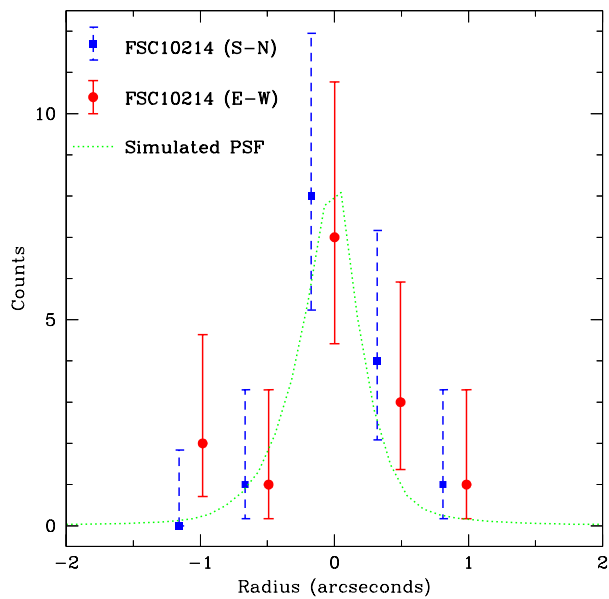}
% \vspace{302pt}
  \caption{The full-band profiles of FSC~10214+4724 (S-N orientation,
    solid dots; E-W orientation, solid squares) compared to the
    on-axis \chandra\ PSF (dotted curve). The $x$-axis corresponds to
    the offset from the {\sc wavdetect}-determined position, and the
    $y$-axis error bars correspond to 1~$\sigma$ uncertainties
    (Gehrels 1986). The \chandra\ PSF was simulated using the {\sc
      ciao} tool {\sc mkpsf} and has been normalised to the peak of
    the S-N orientation profile. Although the signal-to-noise ratio of
    the data is low, the profiles and the half-power radius
    ($\approx$~0\farcs5--0\farcs75) are consistent with those of an
    unresolved source, suggesting a lensing boost in the X-ray
    band of $\simgt$~25; see \S2.2.}
\end{figure}

\section{Discussion}

With a $\approx$~20~ks \chandra\ ACIS-S observation we have shown that
FSC~10214+4724 is comparatively weak at X-ray energies ($L_{\rm
  1.6-6.6~keV}< 3.4\times10^{42}$~erg~s$^{-1}$, $L_{\rm 6.6-26.3~keV}<
6.4\times10^{42}$~erg~s$^{-1}$, and $L_{\rm 1.6-26.3~keV}<
9.8\times10^{42}$~erg~s$^{-1}$ for a lensing boost of $>25$). Previous
studies have suggested that FSC~10214+4724 hosts both a powerful
starburst and a powerful AGN (Goodrich \etal 1996; Granato \etal 1996;
Green \& Rowan-Robinson 1996). In this final section we predict the
expected X-ray emission from both star formation and AGN activity in
FSC~10214+4724 and compare it to the observed X-ray emission. We
also compare the X-ray properties of FSC~10214+4724 to those of
high-redshift \scuba\ galaxies and discuss the X-ray identification of
Compton-thick AGNs at high redshift.

%
%%%%%%%%%%%%%%%%%%%%%%%%%%%%%%%%%%%%%%%%%%%%%%%%%%%%%%%%%%%%%%%%%%%%%%
% FIGURE 3: X-ray versus radio
%%%%%%%%%%%%%%%%%%%%%%%%%%%%%%%%%%%%%%%%%%%%%%%%%%%%%%%%%%%%%%%%%%%%%%
%

\begin{figure}
  \includegraphics[angle=0,width=85mm]{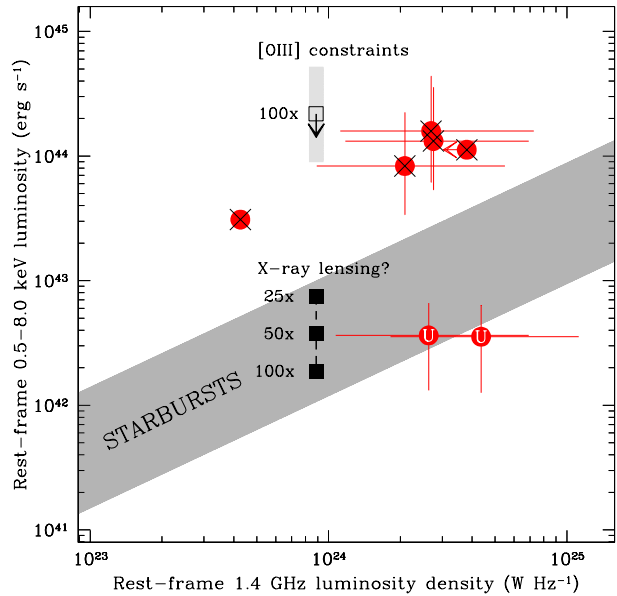}
% \vspace{302pt}
  \caption{Rest-frame 0.5--8.0~keV luminosity versus rest-frame 1.4
    GHz luminosity density for FSC~10214+4724 and a sample of X-ray
    detected \scuba\ galaxies. The filled squares indicate the X-ray
    constraints for FSC~10214+4724 for lensing boosts of 25--100. The
    radio luminosity density of FSC~10214+4724 is calculated assuming
    a lensing boost of 50. The open square indicates the derived upper
    limit on the X-ray emission from the [OIII]$\lambda$5007
    luminosity of FSC~10214+4724 for a lensing boost of 100 (Simpson
    \etal 2004; see \S3.1); the light shaded region shows the variance
    in the $L_{\rm [OIII]}$/$L_{\rm X}$ relationship (Mulchaey \etal
    1994). The filled circles indicate the X-ray detected \scuba\
    galaxies from Alexander \etal (2003b); the crosses indicate
    sources classified as AGNs, and the ``U''s indicate sources with
    unknown classifications but with X-ray properties consistent with
    those of starburst galaxies. The dark shaded region denotes the
    1~$\sigma$ dispersion in the locally determined X-ray-radio
    correlation for star-forming galaxies (see Figure~6 of Shapley,
    Fabbiano, \& Eskridge 2001; Bauer \etal 2002; Ranalli \etal 2003).
    The rest-frame $\simlt$~8~keV emission from FSC~10214+4724 is
    consistent with that expected from star-formation activity; see
    \S3.1.}
\end{figure}

\subsection{The Nature of the X-ray Emission from FSC~10214+4724}

Many studies have shown a correlation between the 1.4~GHz radio
luminosity density and the X-ray luminosity of star-forming galaxies
(e.g.,\ Shapley \etal 2001; Bauer \etal 2002; Ranalli \etal 2003).
Since the radio emission from FSC~10214+4724 is consistent with
star-formation activity (e.g.,\ Lawrence \etal 1993; Rowan-Robinson
\etal 1993; Eisenhardt \etal 1996), we can use this correlation to
predict the expected X-ray emission from star formation. The radio
extent and morphology of FSC~10214+4724 are similar to those found in
the rest-frame ultra-violet, suggesting that the radio emission is
lensed by a factor of \hbox{$\approx$~50--100} (Eisenhardt \etal
1996). In Figure~3 we show the rest-frame 1.4~GHz radio luminosity
density versus the rest-frame 0.5--8.0~keV luminosity for
FSC~10214+4724; the rest-frame 0.5--8.0~keV luminosity was calculated
from the rest-frame 1.6--6.6~keV luminosity (observed soft band)
assuming $\Gamma=1.6$ and is shown for a range of lensing boosts. The
rest-frame \hbox{0.5--8.0~keV} emission from FSC~10214+4724 is
entirely consistent with that expected from star formation for the
range of probable lensing boosts at radio wavelengths. The
X-ray-to-optical flux ratio is also concordant with that expected from
star formation [$\log{({{f_{\rm FB}}\over{f_{\rm I}}})}\simlt-1$; see
\S4.1.1 of Bauer \etal 2004] when appropriate $K$ corrections and
lensing boosts are applied [$\Gamma=1.6$ with lensing boosts of
25--100 in the X-ray band, and the host galaxy templates of Mannucci
\etal (2001) with a lensing boost of 100 in the $I$-band].  These
results imply that the AGN in FSC~10214+4724 is either heavily
obscured or intrinsically weak.

We can estimate the instrinsic luminosity of the AGN in FSC~10214+4724
using the [OIII]$\lambda$5007 luminosity (e.g.,\ Mulchaey \etal 1994;
Bassani \etal 1999). Taking the [OIII]$\lambda$5007 luminosity from
Serjeant \etal (1998), the [OIII]$\lambda$5007 to X-ray correlation of
Mulchaey \etal (1994), and assuming the lensing boost to the
[OIII]$\lambda$5007 emission-line region is $\simgt100$ (Simpson \etal
2004), the predicted rest-frame 0.5--8.0~keV luminosity is
$<2.2\times10^{44}$~erg~s$^{-1}$ (with a variance of
\hbox{$<0.9$--5.2}$\times10^{44}$~erg~s$^{-1}$); see
Figure~3.\footnote{We converted from the 2--10~keV band used by
  Mulchaey \etal (1994) to the 0.5--8.0~keV band assuming
  $\Gamma=2.0$, a typical intrinsic X-ray spectral slope for AGNs.}
These predicted X-ray luminosities are within the range expected for
quasars and would be even higher if the [OIII]$\lambda$5007
emission-line region suffers from reddening (e.g.,\ Elston \etal 1994;
Soifer \etal 1995; cf Serjeant \etal 1998).  Since the lensing boost
to the [OIII]$\lambda$5007 emission-line region is a lower limit,
these constraints should be considered upper limits. However, given
that the [OIII]$\lambda$5007 emission-line region is likely to be more
extended than the central source, the [OIII]$\lambda$5007 emission is
unlikely to be much more magnified than the X-ray emission.

These results suggest that the AGN in FSC~10214+4724 is powerful.
However, the rest-frame 1.6--26.3~keV luminosity is approximately 1--2
orders of magnitude below the constraint estimated from the
[OIII]$\lambda$5007 luminosity. This is significant since rest-frame
$>10$~keV emission is not easily attenuated [e.g.,\ one order of
magnitude of extinction at $>10$~keV requires Compton-thick
obscuration ($N_{\rm H}>1.5\times10^{24}$~cm$^{-2}$); see Appendix~B
in Deluit \& Courvoisier 2003]. Hence, if FSC~10214+4724 hosts a
quasar, as previously suggested, then it must be obscured by
Compton-thick material; these general conclusions are consistent with
those found for other {\it IRAS} galaxies of similar luminosity
(e.g.,\ Iwasawa \etal 2001; Wilman \etal 2003). Under this assumption,
the observed X-ray emission from the AGN would be due to reflection
and scattering and a strong Fe~K$\alpha$ emission line should be
detected.  With only $\approx$~14 X-ray counts, the current X-ray
observations cannot provide good constraints on the presence of
Fe~K$\alpha$; however, a scheduled $\approx$~50~ks \xmm\ observation
(PI: K.~Iwasawa) may be able to place some constraints.

\subsection{\scuba\ Galaxies and Compton-thick Accretion}

The lensing-corrected properties of FSC~10214+4724 are similar to
those of \scuba\ galaxies (see \S1), sources that are probably the
progenitors of massive galaxies (and hence massive black holes) in the
local Universe. The steep X-ray spectral slope and moderately luminous
X-ray emission of FSC~10214+4724 contrasts with the X-ray properties
of the five X-ray detected \scuba\ galaxies classified as AGNs in
Alexander \etal (2003b); however, the X-ray properties of
FSC~10214+4724 are similar to those of the two X-ray detected \scuba\
galaxies classified as unknown (see Figure~3). Since we know that an
AGN is present in FSC~10214+4724 (possibly powerful and Compton
thick), this suggests that further unidentified AGNs may be present in
the \scuba\ galaxy population and the $\approx$~40\% AGN fraction
(Alexander \etal 2003b, 2004) should be considered a lower limit.
Many Compton-thick AGNs may be present in ultra-deep X-ray surveys
(e.g.,\ Fabian \etal 2002). However, the presence of vigorous star
formation may make them difficult to identify on the basis of their
X-ray properties alone.

%We have detected rest-frame 1.6--26.3~keV emission from the strongly
%lensed \hbox{$z=2.285$} ultra-luminous infrared galaxy FSC~10214+4724.
%We estimate the X-ray emission is likely to be lensed by a factor of
%$>25$. We find the rest-frame 1.6--6.6~keV emission is entirely
%consistent with that expected from star formation, suggesting that the
%AGN in FSC~10214+4724 is either heavily obscured or intrinsically
%weak. If an obscured quasar is present in FSC~10214+4724, as
%previously suggested, then it would have to be obscured by Compton
%thick material.

\section*{Acknowledgments}

We acknowledge support from the Royal Society (DMA), PPARC (FEB; CS),
CXC grant GO4-5105X (WNB), NSF CAREER award AST-9983783 (WNB), and
MIUR (COFIN grant 03-02-23; CV). We thank P.~Eisenhardt, D.~Hogg,
K.~Iwasawa, and N.~Trentham for useful discussions.

\label{lastpage}


\begin{thebibliography}{}

\bibitem[]{} 
Abazajian, K., et al.\ 2004, \aj, 128, 502

\bibitem[]{} 
Alexander, D.~M., et al.\ 2003a, \aj, 126, 539 

\bibitem[]{} 
Alexander, D.~M., et al.\ 2003b, \aj, 125, 383 

\bibitem[]{} 
Alexander, D.~M., Smail, I., Bauer, F.~E., Chapman, S.~C., Blain, A.~W., Brandt, W.~N., \& Ivison, R.~J.\ 2004, \nat, submitted

%\bibitem[]{} 
%Almaini, O., et al.\ 2000, \mnras, 315, 325 

\bibitem[]{} 
Bauer, F.~E., Alexander, D.~M., Brandt, W.~N., Hornschemeier, A.~E., Vignali, C., Garmire, G.~P., \& Schneider, D.~P. 2002, \aj, 124, 2351

\bibitem[]{} 
Bauer, F.~E., Alexander, D.~M., Brandt, W.~N., Schneider, D.~P., Treister, E., Hornschemeier, A.~E., \& Garmire, G.~P. 2004, \aj, in press (astro-ph/0408001 )

\bibitem[]{} 
Bassani, L., Dadina, M., Maiolino, R., Salvati, M., Risaliti, G., della Ceca, R., Matt, G., \& Zamorani, G.\ 1999, \apjs, 121, 473

\bibitem[]{} 
Blain, A.~W., Smail, I., Ivison, R.~J., Kneib, J.-P., \& Frayer, D.~T. 2002, Physics Reports, 369, 111

\bibitem[]{} 
Blain, A.~W., Chapman, S.~C., Smail, I., \& Ivison, R.\ 2004, \apj, 611, 725

\bibitem[]{} 
Broadhurst, T.~\& Lehar, J.\ 1995, \apjl, 450, L41 

\bibitem[]{} 
Chapman, S.~C., Blain, A.~W., Ivison, R.~J., Smail, I.~R.\  2003, Nature, 422, 695

\bibitem[]{} 
Chapman, S.~C., Smail, 
I., Windhorst, R., Muxlow, T., \& Ivison, R.~J.\ 2004, \apj, 611, 732 

%\bibitem[]{} 
%Clements, D.~L., van der Werf, P.~P., Krabbe, A., Blietz, M., Genzel, R., \& Ward, M.~J.\ 1993, \mnras, 262, L23 

\bibitem[]{} 
Deluit, S.~\& Courvoisier, T.~J.-L.\ 2003, \aap, 399, 77 

\bibitem[]{} 
Downes, D., Solomon, P.~M., \& Radford, S.~J.~E.\ 1995, \apjl, 453, L65 

\bibitem[]{} 
Ebeling, H., White, D.A., \& Rangarajan, F.V.N.\ 2004, \mnras, submitted

\bibitem[]{} 
Eisenhardt, P.~R., Armus, L., Hogg, D.~W., Soifer, B.~T., Neugebauer, G., \& Werner, M.~W.\ 1996, \apj, 461, 72 

\bibitem[]{} 
Elston, R., McCarthy, P.~J., Eisenhardt, P., Dickinson, M., Spinrad, H., Januzzi, B.~T., \& Maloney, P.\ 1994, \aj, 107, 910 

\bibitem[]{} 
Evans, A.~S., Scoville, N.~Z., Dinshaw, N., Armus, L., Soifer, B.~T., Neugebauer, G., \& Rieke, M.\ 1999, \apj, 518, 145 

\bibitem[]{} 
Fabian, A.~C., Wilman, R.~J., \& Crawford, C.~S.\ 2002, \mnras, 329, L18 

\bibitem[]{} 
Freeman, P.E., Kashyap, V., Rosner, R., \& Lamb, D.Q. 2002, ApJS, 138, 185

\bibitem[]{} 
Garmire, G.~P., Bautz, M.~W., Ford, P.~G., Nousek, J.~A., \& Ricker, G.~R.\ 2003, Proc. SPIE, 4851, 28

\bibitem[]{} 
Gehrels, N. 1986, ApJ, 303, 336

\bibitem[]{} 
George, I.~M., Turner, T.~J., Yaqoob, T., Netzer, H., Laor, A., Mushotzky, R.~F., Nandra, K., \& Takahashi, T.\ 2000, \apj, 531, 52

\bibitem[]{} 
Goodrich, R.~W., Miller, J.~S., Martel, A., Cohen, M.~H., Tran, H.~D., Ogle, P.~M., \& Vermeulen, R.~C.\ 1996, \apjl, 456, L9 

\bibitem[]{} 
Granato, G.~L., Danese, L., \& Franceschini, A.\ 1996, \apjl, 460, L11 

\bibitem[]{} 
Green, S.~M.~\& Rowan-Robinson, M.\ 1996, \mnras, 279, 884 

\bibitem[]{} 
Ivison, R.~J., et al. 2002, \mnras, 337, 1

\bibitem[]{} 
Iwasawa, K.\ 2001, AIP Conf.~Proc.~599: X-ray Astronomy: Stellar Endpoints, AGN, and the Diffuse X-ray Background, 599, 169 

\bibitem[]{} 
Iwasawa, K., Fabian, A.~C., \& Ettori, S.\ 2001, \mnras, 321, L15

%\bibitem[]{} 
%Jannuzi, B.~T., Elston, R., Schmidt, G.~D., Smith, P.~S., \& Stockman, H.~S.\ 1994, \apjl, 429, L49 

\bibitem[]{} 
Lacy, M., Rawlings, S., \& Serjeant, S.\ 1998, \mnras, 299, 1220 

\bibitem[]{} 
Lawrence, A., Rigopoulou, D., Rowan-Robinson, M., McMahon, R.~G., Broadhurst, T., \& Lonsdale, C.~J.\ 1994, \mnras, 266, L41 

\bibitem[]{} 
Lawrence, A., et al.\ 1993, \mnras, 260, 28 

\bibitem[]{} 
Lyons, L. 1991, Data Analysis for Physical Science Students. Cambridge University Press, Cambridge

\bibitem[]{} 
Mannucci, F., Basile, 
F., Poggianti, B.~M., Cimatti, A., Daddi, E., Pozzetti, L., \& Vanzi, L.\ 
2001, \mnras, 326, 745 

\bibitem[]{} 
Mulchaey, J.~S., Koratkar, A., Ward, M.~J., Wilson, A.~S., Whittle, M., Antonucci, R.~R.~J., Kinney, A.~L., \& Hurt, T.\ 1994, \apj, 436, 586 

\bibitem[]{} 
Nandra, K.~\& Pounds, K.~A.\ 1994, \mnras, 268, 405 

%\bibitem[]{} 
%Nandra, K., George, I.~M., Mushotzky, R.~F., Turner, T.~J., \& Yaqoob, T.\ 1997, \apj, 476, 70 

\bibitem[]{} 
Neri, R., et al.\ 2003, \apjl, 597, L113 

\bibitem[]{} 
Nguyen, H.~T., Eisenhardt, P.~R., Werner, M.~W., Goodrich, R., Hogg,
D.~W., Armus, L., Soifer, B.~T., \& Neugebauer, G.\ 1999, \aj, 117,
671

%\bibitem[]{} 
%Norman, C., et al.\ 2002, \apj, 571, 218 

\bibitem[]{} 
Pier, J.~R., Munn, J.~A., Hindsley, R.~B., Hennessy, G.~S., Kent, S.~M., Lupton, R.~H., \& Ivezi{\' c}, {\v Z}.\ 2003, \aj, 125, 1559 

\bibitem[]{} 
Ranalli, 
P., Comastri, A., \& Setti, G.\ 2003, \aap, 399, 39 

\bibitem[]{} 
Rowan-Robinson, M.\ 2000, \mnras, 316, 885

\bibitem[]{} 
Rowan-Robinson, M., et al.\ 1993, \mnras, 261, 513

\bibitem[]{} 
Rowan-Robinson, M., et al.\ 1991, \nat, 351, 719 

%\bibitem[]{} 
%Serjeant, S., Lacy, M., Rawlings, S., King, L.~J., \& Clements, D.~L.\ 1995, \mnras, 276, L31

\bibitem[]{} 
Serjeant, S., Rawlings, S., Lacy, M., McMahon, R.~G., Lawrence, A., Rowan-Robinson, M., \& Mountain, M.\ 1998, \mnras, 298, 321

\bibitem[]{} 
Shapley, A., Fabbiano, G., \& Eskridge, P.~B.\ 2001, \apjs, 137, 139

\bibitem[]{} 
Simpson, C., et al.\ 2004, ApJ, in preparation

\bibitem[]{} 
Smail, I., Ivison, R.~J., Blain, A.~W., \& Kneib, J.-P. 2002, MNRAS, 331, 495

\bibitem[]{} 
Soifer, B.~T., Cohen, J.~G., Armus, L., Matthews, K., Neugebauer, G., \& Oke, J.~B.\ 1995, \apjl, 443, L65 

\bibitem[]{} 
Stark, A.~A., Gammie, C.~F., Wilson, R.~W., Bally, J., Linke, R.~A., Heiles, C., \& Hurwitz, M. 1992, ApJS, 79, 77 

\bibitem[]{} 
Townsley, L.~K., Broos, P.~S., Garmire, G.~P., \& Nousek, J.~A.\ 2000, \apjl, 534, L139 

\bibitem[]{} 
Trentham, N.\ 1995, \mnras, 277, 616 

%\bibitem[]{} 
%Turner, T.~J., George, I.~M., Nandra, K., \& Mushotzky, R.~F.\ 1997, \apjs, 
%113, 23 

\bibitem[]{} 
Weisskopf, M.C., Tananbaum, H.D., Van Speybroeck, L.P., \& O'Dell, S.L.
2000,  Proc. SPIE, 4012, 2 

\bibitem[]{} 
Wilman, R.~J., Fabian, A.~C., Crawford, C.~S., \& Cutri, R.~M.\ 2003, 
\mnras, 338, L19 

%\bibitem[]{} 
%York, D.~G., et al.\ 2000, \aj, 120, 1579 

\end{thebibliography}
\end{document}